\documentclass[12pt,twoside]{article}   %For printing on two sides of page
\usepackage[super,sort,comma]{natbib}
\usepackage{makecell}

\usepackage{fancyhdr}		%Gives headers and footers defined below
\usepackage[section]{placeins}   %
\usepackage{multirow}
\usepackage{graphicx}
\usepackage{float}
\usepackage{amsmath}

\makeatletter \renewcommand\@biblabel[1]{$^{#1}$} \makeatother
\newcommand{\cen}[1]{\begin{center} #1 \end{center}}

\usepackage{hyperref}
\hypersetup{ colorlinks,
    citecolor=blue,
    filecolor=blue,
    linkcolor=blue,
    urlcolor=blue
}

\usepackage{xcolor}
\definecolor{gray}{rgb}{0.6,0.6,0.6}
\definecolor{red}{rgb}{0.85,0,0}
\definecolor{green}{rgb}{0,0.85,0}
\definecolor{blue}{rgb}{0,0,0.85}
\definecolor{beige}{rgb}{0.92,0.87,0.78}
\usepackage[all]{hypcap}    %causes link to figures to go to figure, not caption
\begin{document}

%\cen{\sf {\Large {\bfseries Classification of Multi-phase CT Scans with Deep Learning and Random Sampling }} \\  

\cen{\sf {\Large {\bfseries Phase Recognition in Contrast-Enhanced CT Scans based on Deep Learning and Random Sampling}} \\  
\vspace*{10mm}
Binh T. Dao$^{1,\textcolor{red}{\dag}}$, Thang V. Nguyen$^{1,\textcolor{red}{\dag}}$, Hieu H. Pham$^{1,2,3}$, and Ha Q. Nguyen$^{1,2}$\\
$^{1}$ Smart Health Center, VinBigData JSC, Hanoi, Vietnam \\
$^{2}$ College of Engineering \& Computer Science, VinUniversity, Hanoi, Vietnam\\
$^{3}$ VinUni-Illinois Smart Health Center, Hanoi, Vietnam\\
%Version typeset \today\\
}

% \cen{\sf {\Large {\bfseries Classification of Multi-phase CT Scans with Deep Learning and Random Sampling } \\  
% \vspace*{10mm}
% Author names here} \\
% Author address(es) here
% \vspace{5mm}\\
% Version typeset \today\\
% }

\pagenumbering{roman}
\setcounter{page}{1}
\pagestyle{plain}
$\dag$ These authors contributed equally to this work\\
\indent Corresponding author: \textcolor{blue}{hieu.ph@vinuni.edu.vn} (Hieu H. Pham)\\
% note, probably best not to use a student's e-mail as it won't be valid for
% very long.

\begin{abstract}
\noindent {\bf Purpose:} A fully automated system for interpreting abdominal computed tomography (CT) scans with multiple phases of contrast enhancement requires an accurate classification of the phases. Current approaches to classify the CT phases are commonly based on 3D convolutional neural network (CNN) approaches with high computational complexity and high latency. This work aims at developing and validating a precise, fast multi-phase classifier to recognize three main types of contrast phases in abdominal CT scans.\\
\\
{\bf Methods:} We propose in this study a novel method that uses a random sampling mechanism on top of deep CNNs for the phase recognition of abdominal CT scans of four different phases: non-contrast, arterial, venous, and others. The CNNs work as a slice-wise phase prediction, while the random sampling selects input slices for the CNN models. Afterward, majority voting synthesizes the slice-wise results of the CNNs, to provide the final prediction at scan level.\\
\\
{\bf Results:} Our classifier was trained on 271,426 slices from 830 phase-annotated CT scans, and when combined with majority voting on 30\% of slices randomly chosen from each scan, achieved a mean F1-score of 92.09\% on our internal test set of 358 scans. The proposed method was also evaluated on 2 external test sets: CTPAC-CCRCC (\textit{N} = 242) and LiTS (\textit{N} = 131), which were annotated by our experts. Although a drop in performance has been observed, the model performance remained at a high level of accuracy with a mean F1-score of 76.79\% and 86.94\% on CTPAC-CCRCC and LiTS datasets, respectively. Our experimental results also showed that the proposed method significantly outperformed the state-of-the-art 3D approaches while requiring less computation time for inference.\\
\\
{\bf Conclusions:}  In comparison to state-of-the-art classification methods, the proposed approach shows better accuracy with significantly reduced latency. Our study demonstrates the potential of a precise, fast multi-phase classifier based on a 2D deep learning approach combined with a random sampling method for contrast phase recognition, providing a valuable tool for extracting multi-phase abdomen studies from low veracity, real-world data.

\end{abstract}
% \note{This is a sample note.}

\newpage     %may or may not be needed

% The table of contents is for drafting and refereeing purposes only. Note
% that all links to references, tables and figures can be clicked on and
% returned to calling point using cmd[ on a Mac using Preview or some
% equivalent on PCs (see View - go to on whatever reader).
\tableofcontents

\newpage

\setlength{\baselineskip}{0.7cm}      %double spacing		

\pagenumbering{arabic}
\setcounter{page}{1}
\pagestyle{fancy}
\section{Introduction}

Contrast enhancement in computed tomography (CT) scans, especially of the abdomen, is crucial for successful lesion diagnosis~\cite{brancatelli2001focal, bronstein2004detection}. Certain types of lesions can only be observed on the CT scans taken after the injection of contrast agents into the blood veins. The contrast enhancement process generally consists of three main phases~\cite{smithuis_2014} as follows.
\begin{itemize}
    \item Non-Contrast: the CT scan is acquired  without injection of any contrast agents;
    \item Arterial: the CT scan is acquired 35-40  seconds  after the bolus injection, which can help with identifying the hepatocellular carcinoma (HCC), the focal nodular hyperplasia (FNH), and the adenoma in the liver;
    \item Venous: the CT scan is acquired 70-80 seconds after the bolus injection, in which the liver parenchyma is enhanced through the blood supply by the portal vein, highlighting hypovascular liver lesions.
\end{itemize}

Machine learning algorithms over the last decades have achieved a great success in the interpretation and diagnosis of medical imaging data~\cite{litjens2017survey}, including the automatic detection of liver lesions in contrast-enhanced CT images~\cite{yasaka2018deep,wang2018classification,yoshinobu2020deep,gao2021deep}. For instance, a multi-phase analysis of abdominal CT scans~\cite{nayak2019computer} was performed to detect cirrhosis and HCC liver. To obtain a robust performance, however, such an algorithm often requires training from a large-scale dataset of patient’s preoperative multi-phase CT scans and clinical features~\cite{gao2021deep}. Hence, a reliable method is needed for the collection and annotation of imaging data of abdominal structures~\cite{park2020annotated}. The data mining process usually starts with accessing and collecting retrospective medical imaging data through the Picture Archiving and Communication Systems (PACS). Unfortunately, the current generation of PACS systems does not support the curation of large-scale, multi-phase contrast-enhanced CT datasets. The key obstacle is that DICOM tags related to series description (\textit{e.g.}, non-contrast, arterial, or venous) are manually input, non-standardized, and often incomplete~\cite{harvey2019standardised}. These limitation leads to the impossibility to automatically categorize medical image data based solely on their DICOM metatags, as around 15\% of all studies were mislabeled due to human factor~\cite{gueld2002quality}. As a result, these datasets often rely on physicians for manual re-annotation of CT scans, typically expensive and time-consuming.

In addition, a typical machine learning-based computer-aided diagnosis (CAD) system for interpreting abdominal CT scans is often trained images from a specific phase, or stacks of images from different phases in some fixed order~\cite{sun2017automatic}. Consequently, in the deployment scenario, it is essential that the system knows precisely which phase each CT scan (series) belongs to, so that the right phase scans can be fed to models. This leads to a dire need for a phase identification module for abdominal CT series.\\
\\

Several approaches~\cite{zhou2019ct,tang2020contrast} have been proposed to identify multiple phases from CT scans. For instance, Zhou \textit{et al}.~\cite{zhou2019ct} focuses on volumetric characteristics of the scan, resizes the entire scan to a 3D block of size 32$\times$ 128$\times$128 by interpolation, then feeds them through a 3D CNN. In the original paper, the proposed model is trained with 43,000 scans, which is a significant amount of data to be acquired, and this sparks the question of whether this method would perform well on our smaller dataset. Another work by Tang \textit{et al}.~\cite{tang2020contrast} suggests using a generative adversarial network on each slice instead of the whole 3D scan. The authors state that CD-GAN is trained for a period of approximately 36 hours on an NVIDIA 2080Ti GPU with 11G memory, showing that this method is too computationally expensive. 

Different from the previous approaches~\cite{zhou2019ct,tang2020contrast}, we aim to develop a fast, highly accurate deep learning system for recognizing phases from CT scans. Specifically, we propose an efficient strategy solely based on a 2D representation of the slices. The proposed system consists of two main stages: (1) \textbf{Random sampling} that randomly picks \textit{R}\% of slices from the input CT scan and inputs to the deep learning model. Here, \textit{R\%} denotes the percentage of slices selected from the input scan; and (2) \textbf{Slice-level prediction} that identifies phases of each chosen slice then uses majority voting to conclude the phase of the given scan. Our experimental results on internal and external (\textit{i.e.}, CTPAC-CCRCC~\cite{kalayci2020pronetview}, LiTS~\cite{bilic2019liver}) datasets showed that the proposed method significantly outperforms the state-of-the-art 3D approaches while requiring less computation time for inference.

To summarize, the main contributions of this work are the following:

$\bullet$ We develop and evaluate a novel deep learning system for the recognition of multi-phase in contrast-enhanced CT scans. The proposed system exploits a random sampler to reduce the reduce computational time of the input examples. Majority voting is used to boost the final prediction of the system. Our extensive experiments show that the proposed approach surpasses previous state-of-the-art 3D approaches in terms of both accuracy and inference time.  The proposed deep learning system can be easily reused or finetuned, therefore potential benefits for several applications in clinical settings. 

$\bullet$  The imaging dataset used in this study will be shared on our project website at \url{https://vindr.ai/datasets/abdomen-phases}, while the codes will be published  at \url{https://github.com/vinbigdata-medical/abdomen-phases}.  To the best of our knowledge, this is the biggest annotated dataset  for the recognition of multi-phase in contrast-enhanced CT scans.

\section{Proposed Approach}
\subsection{Overview of approach}
Our main goal in this study is to develop and evaluate a fast, accurate deep learning system for the recognition of multi-phase in contrast-enhanced CT scans. To this end, we first randomly sample $R$\% of the slices from the whole original scan. Each of these chosen slices is then passed through a CNN model, which was trained to output the phase classification at the slice level. Finally, the scan-level prediction is predicted by a majority voting of the results obtained in the previous step. The proposed scheme for the phase recognition of abdominal CT scans is illustrated in Figure 2.%\autoref{ModelPipeline}.

%\begin{figure}
%   \begin{center}
%   \includegraphics[width=16cm]{Pipeline.png}
 %  \captionv{16}{}{The overall pipeline for the phase prediction from abdominal CT scans. The slices are passed sequentially through a single CNN model. Using predicted labels produced by the CNNs, a majority voting is performed to boost system performance.
 % \label{ModelPipeline}
  %  }  %note label inside caption
 %   \end{center}
%\end{figure}  

\subsection{Data collection and annotation}
To develop deep learning algorithms for contrast phase recognition, we built an internal dataset of abdominal CT scans. The construction of this dataset was divided into three main steps: (1) data collection, (2) data de-identification, and (3) data annotation.\\
\\
\textbf{Data collection}:   A total 265 abdominal studies comprising 1,188 CT scans in the Digital Imaging and Communications in Medicine (DICOM) format were retrospectively randomly selected from the PACS databases of the Hospital 108 and the Hanoi Medical University Hospital -- two major hospitals in Vietnam -- within the period from 2015 until 2020. The ethical clearance of this study was approved by the Institutional Review Board (IRB) of each hospital before any data processing steps. The need for obtaining patient consent was waived because these studies did not impact clinical care.

Data characteristics, including patient demographics and the prevalence of each contrast-phase class, are summarized in Table 1. A general statistics of the slice and scan distribution for each class is featured in Figure 3. The distribution of the CT scanner models and their manufacturers are shown in Figure 4. There are 6 different levels of slice thickness in the entire dataset, whose distribution is captured in Table 2. Additionally, the number of slices per scan ranges from 30 to 2,350 with a mean of 281. The distribution of the number of slices per scan is illustrated in Figure 5.\\
\\
\textbf{Data de-identification}: To protect patient's privacy, all personally identifiable information associated with the DICOM images has been removed. Specifically, a Python script was written to remove all DICOM tags of protected health information (PHI) such as patient's name, patient's date of birth, patient ID, etc. We only retained a limited number of DICOM attributes that are necessary for processing raw images. The full list of these DICOM attributes is provided in Supplement~\ref{list-dicom-tag}, Table 9.\\
\\
\textbf{Data annotation}: The dataset was labeled for a total of 4 contrast-phase classes: (1) Non-contrast, (2) Venous, (3) Arterial, and (4) Others. Here the Others category refers to all scans that cannot be correctly classified as either of 3 phases Non-contrast, Venous, and Arterial. They may include scans of the delay phase or scans that belong to a transitional state between 2 phases. To annotate the imaging data, we designed and built a web-based labeling framework called VinDr Lab (\url{https://vindr.ai/vindr-lab})~\cite{VinDr-Lab}. Two radiologists were hired to remotely annotate the data. Once the labeling has been completed, the labels were exported in Comma-Separated Values (CSV) format and will be used for training deep learning algorithms.

\noindent In total, 265 studies have been annotated. The whole dataset was then divided into training and validation set by a ratio of 70\%/30\% accordingly. Since each study usually contains multiple scans of the same patient, the train-test split was stratified by the study level to avoid data leakage. As a result, our training set consists of 271,426 slices from 830 scans (186 studies), while our validation set contains 121,134 slices from 358 scans (79 studies). \\
\\
\textbf{Data Records}: To encourage new advances in this research direction, we will make the dataset freely accessible via our project website at \url{https://vindr.ai/datasets/abdomen-phases}. Specifically, all imaging data and the corresponding ground truth labels for the training and validation set will be provided. The images are organized into two folders, one for training and the other one for validation in which each image has a unique, anonymous identifier. \\

\subsection{Model development}

This section describes in detail our model development method. We exploit state-of-the-art, high-performing deep CNN architectures for the task of recognizing multi-phase in contrast-enhanced CT scans. We describe our network architecture choice and training methodology as the following.\\
\subsubsection{Network architecture}

A set of state-of-the-art deep CNN models has been deployed and evaluated on the collected dataset, including ResNet-18~\cite{he2016deep}, ResNet-34~\cite{he2016deep}, SEResNet-18, ResNext-50~\cite{xie2017aggregated}, EfficientNet-B0~\cite{tan2019efficientnet}, EfficientNet-B2~\cite{tan2019efficientnet},
GhostNet~\cite{han2020ghostnet}, and CD-GAN~\cite{tang2019contrast}. These deep networks are well-known to be effective for image recognition tasks. Each network accepts a CT scan as input and predicts the corresponding contrast phase label. For implementation, we followed the same instructions and recommendations from the original papers~\cite{he2016deep,xie2017aggregated,tan2019efficientnet,han2020ghostnet,tang2019contrast}. We considered EfficientNet~\cite{tan2019efficientnet} model as our main network architecture choice due to the high-level of accuracy and efficiency of this architecture compared to previous deep CNNs. Details of the EfficientNet~\cite{tan2019efficientnet} architecture is provided in Supplement~\ref{efficientnet}.  

\subsubsection{Training methodology}

In the training stage, all images were fed into the networks with a size of 224$\times$224 pixels. Input images extracted from raw DICOM files were, first, converted to standard Hounsfield units (HU), using Rescale Slope and Rescale Intercept from DICOM headers. Afterward, we applied HU window with the window center of 50 and the window width of 400 to the image. During the training process, we used Adam optimizer~\cite{kingma2014adam} with an initial learning rate of $10^{-2}$ and cosine annealing scheduler~\cite{loshchilov2016sgdr} with linear warm up~\cite{ma2019adequacy}. Each network was trained end-to-end for 15 epochs. To this end, we minimized the binary cross-entropy loss function between the ground-truth labels and the predicted label by the network over the training samples. The proposed deep network was implemented in Python using Pytorch version 1.7.1 (\url{https://pytorch.org/}). All experiments were conducted on a Ubuntu 18.04 machine with a single NVIDIA Geforce RTX 2080 Ti with 11GB memory.

% Input to our model are 2D input image read from raw DICOM file. Firstly, the raw image was converted to the standard Hounsfield Units (HU), using Rescale Slope and Rescale Intercept extracted from metadata of each file. Then we apply HU windows with window center of 50 and window width of 400 to the image. This highlights the liver and kidney where contrast material passes through.   
% We, then, trained the slice-level classifier using the EfficientNet-b2 \cite{tan2019efficientnet} architecture on a set of 271,426 slices from 830 scans of 186 studies. The model was trained in PyTorch using the Adam optimizer with an initial learning rate of $10^{-2}$ and cosine annealing scheduler with linear warm up. All slices were resized to 128$\times$128 before being fed to the model. The model was validated on a test set of 122,641 slices from 396 scans of 79 studies. 

\section{Experiments and Results}

\subsection{Experimental setup}

\textbf{Internal validation}: Extensive experiments were conducted to evaluate the performance of the proposed method. Specifically, we first evaluated the slice-wise classification performance of trained CNN models (i.e., ResNet-18~\cite{he2016deep}, ResNet-34~\cite{he2016deep}, SEResNet-18, ResNext-50~\cite{xie2017aggregated}, EfficientNet-B0~\cite{tan2019efficientnet}, EfficientNet-B2~\cite{tan2019efficientnet},
GhostNet~\cite{han2020ghostnet}, and CD-GAN~\cite{tang2019contrast}) on the validation set of 121,134 slices. Next, we reported the classification performance of the best performing network at the scan-wise level by applying the majority voting on \textit{R}\% of the slices selected from each scan. We experimented with \textit{R} ranging from 1 to 20 at an interval of 5, then 20 to 100 at an interval of 10. Finally, to compare the proposed 2D approach with previous 3D state-of-the-art approaches, we re-implemented two-phase recognition approaches on CT scans including 3DSE~\cite{zhou2019ct} and CD-GAN~\cite{tang2019contrast}. These approaches were trained on the training dataset using the same hyperparameters setting as described in the original papers~\cite{zhou2019ct,tang2019contrast}. We also measured the average inference time (second) per scan for each approach and compared it with our proposed 2D method.\\
\\
\textbf{External validation}: To verify the generalization ability of the proposed deep learning model, we evaluated it on two external datasets, including LiTS~\cite{bilic2019liver} and CPTAC-CCRCC~\cite{kalayci2020pronetview}. The LiTS~\cite{bilic2019liver} dataset contains 131 CT scans in the training set and 70 CT scans in the test set. It was original developed for the development of liver segmentation methods. The CPTAC-CCRCC~\cite{kalayci2020pronetview} was introduced by the National Cancer Institute's Clinical Proteomic Tumor Analysis Consortium (CPTAC) and developed for investigating the clear cell renal cell carcinoma (CCRCC).  We utilized the imaging data for CCRCC tumors, containing 242 CT scans for our external testing. However, the original aim of these datasets did not match the purpose of this study. As a result, there were no target labels for this dataset. Our radiologist team, therefore, classified scans from these datasets into 4 phase categories. As a result, the LiTS~\cite{bilic2019liver} dataset has 8 scans of the Arterial phase, and 123 scans of the Venous phase. Meanwhile, the CPTAC-CCRCC~\cite{kalayci2020pronetview} contains 57, 69, 53, and 63 scans from 4 categories Non-Contrast, Venous, Arterial, and Others, respectively.

\subsection{Evaluation metrics}

We report the classification performance using mean accuracy, macro-average precision precision, macro-average precision recall, macro-average precision and macro-average F1-score. These performance indicators are defined as follows.
\begin{align*}
    \text{Accuracy} &= \frac{\text{TP} + \text{TN}}{\text{TP} + \text{FP} + \text{TN} + \text{FN}}, \\
    \text{Precision} &= \frac{\text{TP}}{\text{TP} + \text{FP}},\\
    \text{Recall} &= \frac{\text{TP}}{\text{TP} + \text{FN}},\\
    \text{F1} &= 2 \times \frac{\text{Precision} \times \text{Recall}}{\text{Precision} + \text{Recall}}.
\end{align*}
Here, TP, FP and FN are the number of True Positive, False Positive and False Negative samples accordingly.

\subsection{Experimental results}

\subsubsection{Model performance on internal test set}

Table 3 summarizes quantitative results for several state-of-the-art CNN classification models on the internal test set of 121,134 slices. Note that, while training and benchmarking those models, we chose to fix the input image size to be 128$\times$128 for the sake of saving computations. It can be seen that EfficientNet-B2 achieved the best performance with a macro-averaged recall of 85.92\%, a macro-averaged precision of 84.70\%, and a macro-averaged F1-score of 85.26\%. This architecture was then selected to conduct all the remaining experiments.\\

%\begin{table}[htbp]
%\centering
% \captionv{14}{}{Experimental results across CNN models on the slice-level evaluation when trained with input images of size 128$\times$128. The best F1-score is in \textbf{bold}.\label{SliceLevelQualitative}}%
 % {\begin{tabular}{|c|c|c|c|c|}\hline
 % \bfseries Network architecture & \bfseries Accuracy & \bfseries Precision & \bfseries Recall & \bfseries F1-score\\
  %\hline
 % ResNet-18~\cite{he2016deep} & 0.8964 & 0.8249 & 0.8152 & 0.8198\\
 % \hline
 % ResNet-34~\cite{he2016deep} & 0.9095 & 0.8456 & 0.8172 & 0.8271\\
 % \hline
 % SEResNet-18\cite{hu2018squeeze} & 0.8957 & 0.8192 & 0.8095 & 0.8141\\
 % \hline
 % ResNext-50~\cite{xie2017aggregated} & 0.9159 & 0.8515 & 0.8444 & 0.8475\\
 % \hline
 % EfficientNet-B0~\cite{tan2019efficientnet} & 0.9229 & 0.8486 & 0.8483 & 0.8484\\
 % \hline
 % \textbf{EfficientNet-B2}~\cite{tan2019efficientnet} & 0.9215 & 0.8470 & 0.8592 & \textbf{0.8526}\\
 % \hline
 % GhostNet~\cite{han2020ghostnet} & 0.9151 & 0.8397 & 0.8398 & 0.8397 \\
 % \hline 
 % CD-GAN~\cite{tang2019contrast} & 0.8979 & 0.8093 & 0.8526 & 0.8219 \\ 
 % \hline
 % \end{tabular}}
 %\vspace{10pt}
%\end{table}

We further investigated the impact of different image input sizes on the performance of the picked model, \textit{i.e.} EfficientNet-B2, as shown in Table 4. We observed that using the input images with a size of 224$\times$224 for training gave us the best result: a macro-averaged accuracy of 93.51\%; a macro-averaged recall of 85.46\%; a macro-averaged precision of 87.45\% and a macro-averaged F1-score of 86.43\%. In addition, training the model with the 224$\times$224 images only took 10 minutes for each epoch instead of 60 minutes when using the input images of size 512$\times$512.

The classification performance of  EfficientNet-B2 on our test set is shown in Table 5. We computed the precision, recall, and F1-score for both the slice and scan levels along with their 95\% confidence interval (CI) using bootstrapping over 5,000 resamples of the test set. Our proposed model achieved a mean F1-score of  0.8643 (95\% CI 0.8602, 0.8664) for the slice-level prediction and a mean F1-score of 0.9209 (95\% CI 0.9033, 0.9374) for the scan-level prediction with majority voting on 30\% of the total slices. We observed that the reported performance remained consistent between our three main classes: Non-Contrast, Venous, and Arterial, while there was a visible gap between Others and the rest. Additionally, the ROC (Receiver Operating Characteristic) curves of the proposed model for the four classes are plotted in Figure 6 along with their corresponding AUC (Area Under the ROC Curve) scores on the slice-level test set. Unlike F1-score, AUC is a threshold-independent metric. Nevertheless, it can be seen that the AUC scores reported in Figure 6 are strongly correlated with the slice-wise F1-scores given in Table 5.

The effect of using different values of \textit{R} when performing random sampling with the majority voting for scan-level prediction is illustrated in Figure 7. It can be clearly seen that the macro-average F1-score increased as \textit{R} (the percentage of slices in each scan to be selected randomly for inference) approaches 20\%, and leveled off as \textit{R} increased. By applying $R = 30$, we observed a 5.66\% increase in macro-averaged F1-score compared to the slice-wise performance. \\

%\begin{figure}[ht]
%   \begin{center}
%   \includegraphics[width=12cm]{Random_sampling_CI.png}
%   \captionv{14}{}{Scan-wise performance (mean F1-score) of the trained EfficientNet-B2 on the internal test set is plotted against the percentage $R$ of randomly sampled slices per scan used in majority voting. The shadow strip depicts the 95\% confidence intervals of these F1-scores. \label{95CI}}
%    \end{center}
%\end{figure}

\subsubsection{Comparison to state-of-the-art} 

To demonstrate the effectiveness of the proposed 2D approach, we compared our result with recent state-of-the-art methods~\cite{zhou2019ct,tang2019contrast} for the recognition of multi-phase in contrast-enhanced CT scans. To this end, we reproduced the 3DSE by Zhou \textit{et al}.~\cite{zhou2019ct} and the CD-GAN~\cite{tang2019contrast} by Tang \textit{et al.} and reported their performance of these approaches using F1-score on the test set. For a fair comparison, we applied the same training methodologies and hyper-parameter settings as reported in the original papers~\cite{zhou2019ct,tang2019contrast}. In particular, the input image size to the model was fixed to 128$\times$128 while comparing to CD-GAN. The experimental results are provided in Table 6. We found that the proposed 2D approach significantly surpassed the previous state-of-the-art approaches (an improvement of 6.09\% compared to the 3DSE~\cite{zhou2019ct} and 3.07\% compared to CD-GAN~\cite{tang2019contrast}), while required less time for inference.\\
\\
%\begin{table}[htbp]
%{\captionv{16}{}{Comparison to state-of-the-art approaches. Best results are in \textbf{bold}.} \label{ScanLevelQualitative}}%
%\centering
 % The first argument is the label.
 % The caption goes in the second argument, and the table contents
 % go in the third argument.
% \small{ 
 %{\renewcommand{\arraystretch}{1.5}% for the vertical padding
 % {\begin{tabular}{c|c|c|c|c|c}\hline
%   \bfseries Model Architecture & \bfseries F1-score & \bfseries Inference time (s)\\
%  \hline
 %  & \textbf{Method} & \textbf{Precision} & \textbf{Recall} & \textbf{F1-score} & \textbf{Infer. time (s)}\\
 % \hline
%  \parbox[t]{8mm}{\multirow{2}{*}{\rotatebox[origin=c]{90}{\textbf{\makecell{Scan\\Wise}}}}} & \textbf{\scriptsize{EfficientNet-B2 + sampling (ours)}} & \textbf{0.9209} & \textbf{0.9220} & \textbf{0.9209} & \textbf{6.87}$\times\textbf{10}^{\textbf{-4}}$\\
 %                       & \scriptsize{3DSENet}~\cite{zhou2019ct} & 0.8288 & 0.9092 & 0.8600 & 2.12 $\times 10^{-3}$\\
                        
 % \hline\hline
  %\parbox[t]{8mm}{\multirow{2}{*}{\rotatebox[origin=c]{90}{\textbf{\makecell{Slice\\Wise}}}}} & \textbf{\scriptsize{EfficientNet-B2 (ours)}} & \textbf{0.8470} & \textbf{0.8592} & \textbf{0.8526} & 8.98$\times 10^{-5}$\\
  %                  & \scriptsize{CD-GAN}~\cite{tang2019contrast} & 0.8093 & 0.8526 & 0.8219 & 1.60$\times 10^{-5}$\\
 % \hline
  %\hline
  %\end{tabular}}
 % }}
%\vspace{10pt}
%\end{table}

\subsubsection{Model performance on external test set}

We reported in Table 7 the experimental results on two external test sets LiTS and CPTAC-CCRCC. The averaged F1-scored on the LiTS was 86.94\%, while the averaged F1-score on the CPTAC-CCRCC was 76.79\%.  We found that the proposed method suffered from covariate shift, however, it still remains at a high level of F1-score.

%\begin{table}
%\centering
%{\captionv{14}{}{Across-class quantitative results on the external datasets.} \label{ExternalPerformance}}
 % {\begin{tabular}{c|c|c|c|c|c}\hline
%   \bfseries Model Architecture & \bfseries F1-score & \bfseries Inference time (s)\\
%  \hline
 %  & \textbf{Categories} & \textbf{Precision} & \textbf{Recall} & \textbf{F1-score} & \textbf{\# Samples}\\
 % \hline
 % \parbox[t]{8mm}{\multirow{4}{*}{\rotatebox[origin=c]{90}{\textbf{\makecell{LiTS}}}}} & Non-Contrast & N/A & N/A & N/A & 0 \\
 %                       & Venous & 0.9868 & 0.9763 & 0.9804 & 124\\
 %                       & Arterial & 0.7312 & 0.7987 & 0.7584 & 7\\
 %                       & Others & N/A & N/A & N/A &  0\\

  %\hline\hline
  %\parbox[t]{8mm}{\multirow{4}{*}{\rotatebox[origin=c]{90}{\textbf{\makecell{CPTAC-\\CCRCC}}}}} & Non-Contrast & 0.7728 & 0.9140 & 0.8374 & 57\\
    %                            & Venous & 0.7018 & 0.8833 & 0.7829 & 69\\
   %                             & Arterial & 0.9077 & 0.8191 & 0.8609 & 53\\
   %                             & Others & 0.7688 & 0.4858 & 0.5905 & 63\\
  %\hline
  %\hline
  %\end{tabular}}
  %\vspace{10pt}
%\end{table}

\section{Discussion}

\subsection{Key findings} 

The phase recognition is important for medical imaging data collection and the deployment of machine learning models in practice. From the clinical perspective, a method for fast and precise recognition of CT phases can effectively aid in diagnosis of abdominal pathologies~\cite{guite2013computed}. By training a set of strong deep CNN models on a large-scale, annotated dataset, we built an automated system that is able to accurately recognize contrast multi-phases from CT scans. In particular, we empirically showed that a major improvement has been achieved, in terms of F1-score and inference time by applying the proposed random sampling and majority voting. Compared to previous state-of-the-art 3D approaches, our model showed 30 times improved in inference time and nearly 6\% improvement on F1-score on our dataset. \\

%\subsection{Limitations}
%\begin{figure}[ht]
%   \begin{center}
 %  \includegraphics[width=10cm]{Discussion.png}
 %  \captionv{12}{}{Aortic area images from scans of 4 different patients demonstrate the visual variance between images of the same categories. \label{Discussion}}
 %   \end{center}
%\end{figure}

Although a highly accurate performance has been achieved across three classes: Non-Contrast, Arterial and Venous, we acknowledge that the proposed method reveals some limitations. To make a correct classification of contrast phases, experts often rely on the multiple slices containing arteries, veins, and parenchyma. However, in our method, slices from each scan are predicted independently, without incorporating information from other regions. In addition, since contrast materials are absorbed differently for each individual, slices of the same regions from 2 different phases, such as the Arterial and Venous phases, could have similar brightness in the arteries. An example is demonstrated in Figure 8: there is a clear brightness difference between the 2 images in the Arterial phase and the top left Arterial image resembles slices from the Venous phase. For this reason, our slice-level predictions are prone to errors. Another challenge has related to the nature of our dataset. Our samples vary in scan range, some abdominal CT scans can include neck or thighs where contrast material does not pass through, making these slices indifferentiable across our 4 classes. Therefore, our model is likely to produce a false prediction on these slices, which contributes negatively to the performance of the scan level prediction.

The low generalizability of deep learning-based diagnostic systems~\cite{therrien2018role,liang2020generalizability,willemink2020preparing,nadeem2021generalizability} to datasets and scanners beyond the ones they have been trained with, has been limiting the use of such methods in real-world clinical settings. We showed that the proposed deep learning method was successfully generalized to two different datasets from other hospital sites, for which each with a different CT scanner.

\subsection{Future works}
There are several possible mechanisms to improve our current method. The most promising direction is to eliminate non-affected contrast-enhanced regions of a scan, such as pelvis areas. This would improve slice-level prediction since the model is forced to learn and predict images with clearer features. In addition, due to the performance drop in ``Others'', future work includes applying techniques for reducing the impact of imbalanced data. For example, weighted BCE losses~\cite{ho2019real}, which directly penalized the probabilistic false positives, can be used. We also plan to experiment with training and testing the proposed method on the coronal and sagittal projections of the CT scans, so that each input image could contain all necessary components: arteries, veins, and parenchyma, which are used to identify the correct contrast phases. Moreover, we would conduct additional experiments, incorporating training procedure refinements~\cite{he2019bag} such as data augmentation methods to further improve the generalization of our method. Lastly, it is worth investigating more sophisticated methods for sampling and synthesizing the slice-level predictions, such as the multi-instance learning paradigm~\cite{Zhennan20216}, rather than the straightforward random sampling and majority voting strategies discussed in the paper.

\section{Conclusion}

In this study, we developed a 2D deep learning-based approach for the recognition of contrast phases in abdominal CT scans. We adopted a random sampling strategy to improve the classification performance and to reduce inference time. The introduction of a random sampling mechanism helps avoid training and inferring on 3D data, which are usually much more costly, while still attaining impressive performances. Extensive experimental results on both the internal and external datasets have demonstrated that the proposed approach significantly outperformed previous state-of-the-art 3D approaches. 

\section{Acknowledgements}
This work was supported by VinBigData JSC. We are especially thankful to all of our collaborators, including radiologists, physicians, and technicians, who participated in the data collection and labeling process.

\section{Conflict of Interest} 

The authors have no conflict of interest to disclose.

% % following only if there is an appendix
% \section*{Appendix}
% \addcontentsline{toc}{section}{\numberline{}Appendix}
% Appendix text goes here if needed.

\section*{References}
\addcontentsline{toc}{section}{\numberline{}References}
\vspace*{-20mm}

% Following assumes you are using bibtex. However, for submission to the
% journal you MUST explicitly INCLUDE THE REFERENCES IN THE TEX FILE. 
% In that case you need the following

% \begin{thebibliography}{10}
% insert the .bbl file generated by bibtex here
	%This will be a series of entries from your .bib file formatted
	%something like
	%\bibitem{Me09}
        %{I.~Meijsing, B.~W.~Raaymakers, A.~J.~E.~Raaijmakers \it et al.},
        %\newblock {Dosimetry for the MRI accelerator: the impact of a 
	%magnetic field on the response of a Farmer NE2571 ionization chamber},
        %\newblock Phys. Med. Biol. {\bf 54}, 2993 -- 3002 (2009).

% \end{thebibliography}

% The following is when using bibtex and picks up the example.bib file

%\bibliography{Explicit address of .bib file}
\bibliography{./example}      %example.bib is on the same directory
% above points to where we find the master reference list
% and also causes the bibliography to be printed

% When creating your bibliography you should run bibtex on your local
% computer after running pdflatex on your .tex file. bibtex will
% generate a .bbl file.
% Copy the contents of this .bbl file into your main latex document,
% replacing the "\bibliography" command which was pointing at your .bib file.

% following defines style of .bbl file 

%\bibliographystyle{explicit relative path to medphy.bst}
\bibliographystyle{./medphy.bst}    %if this is installed on your system,
				    %it is not essential to have the    ./

% Note that you need to typeset once, then run bibtex, then typeset another
% two times to get the references working properly.

\section{Supplementary Materials}
\subsection{EfficientNet architecture \label{efficientnet}}
This section provides a detailed description of EfficientNet~\cite{tan2019efficientnet}, which is our main network architecture choice in this work. In 2019, Mingxing Tan \textit{et al}. proposed EfficientNet~\cite{tan2019efficientnet} that achieve state-of-the-art performance on top-1 accuracy on ImageNet~\cite{deng2009imagenet} dataset by the time their model architecture was first released. In addition to the improvement of accuracy, the author also claimed to have significantly improved the model efficiency. One of the main contribution of Mingxing Tan \textit{et al}.~\cite{tan2019efficientnet} was the compound scaling method demonstrated in Figure 9. \textbf{Width} simply means how wide the network is, in terms of the number of channels in a Conv layer. Wider networks can capture more fine-grained features, keeping the model small. However, the networks will saturate quickly with larger width as in-depth scaling. \textbf{Depth} means how many layers are stacked into the network. Scaling by depth dimension enables the model to capture richer and more complex features. Yet, as the network gets deeper, the performance will be affected by gradient vanishing. \textbf{Resolution} means the resolution of the input image to the network. The higher the resolution, the more features are captured during training. Nevertheless, this does not scale linearly, and quickly diminish with the larger image size. The proposed compound scaling is introduced to solve these problems. The method is further illustrated in the formula below. \\

%\begin{table}
%  {\captionv{14}{}{EficientNet-B0 baseline network architecture} \label{NetworkArchitecture}}%
%\centering
%  {\begin{tabular}{|c|c|c|c|c|}\hline
%   \textbf{Stage i} &  \textbf{Operator} &  \textbf{Resolution} & \textbf{\#Channels} & \textbf{\#Layers}\\
%  \hline
%  1 & Conv3x3 & 224$\times$224 & 32 & 1\\
%  \hline
%  2 & MBConv1, k3$\times$3 & 112$\times$112 & 16 & 1\\
%  \hline
%  3 & MBConv6, k3$\times$3 & 112$\times$112 & 24 & 2\\
%  \hline
%  4 & MBConv6, k5$\times$5 & 56$\times$56 & 40 & 2\\
%  \hline
%  5 & MBConv6, k3$\times$3 & 28$\times$28 & 80 & 3\\
%  \hline
%  6 & MBConv6, k5$\times$5 & 14$\times$14 & 112 & 3\\
%  \hline
%  7 & MBConv6, k5$\times$5 & 14$\times$14 & 192 & 4\\
%  \hline 
%  8 & MBConv6, k3$\times$3 & 7$\times$7 & 320 & 1 \\
%  \hline 
%  9 & Conv1x1 \& Pooling \& FC & 7$\times$7 & 1280 & 1 \\ 
%  \hline
 % \end{tabular}}
 % \vspace{10pt}
%\end{table}

\newenvironment{tightcenter}{%
  \setlength\topsep{0pt}
  \setlength\parskip{0pt}
  \begin{center}
}{%
  \end{center}
}
\begin{tightcenter}
depth: $d = \alpha ^{\phi }$ \\
width: $w = \beta ^{\phi }$ \\ 
resolution: $r = \gamma ^{\phi }$ \\
s.t. $\alpha \cdot \beta^{2} \cdot \gamma^{2} \approx 2$ \\
$\alpha \geq 1, \beta \geq 2, \gamma \geq 1$ \\

\end{tightcenter}

where $\alpha$, $\beta$, and $\gamma$ are factors that can be determined using grid search. Here $\phi$ controls how many more resources are available for model scaling, whereas $\alpha$, $\beta$, and $\gamma$ specify how to distribute these extra resources to the network width, depth, and resolution.
In addition to the novel scaling method, the authors introduced a baseline EfficientNet-B0, by doing a Neural Architecture Search that optimized both accuracy and floating-point operations per second (FLOPS). The architecture of the baseline is depicted in Table 8. The main building block of the network is the mobile inverted bottleneck MBConv added with squeeze-and-excitation optimization. By fixing $\alpha$, $\beta$, $\gamma$ and experiment with different values of $\phi$, we get EfficientNet B1-B7.
\end{document}